\documentclass[conference]{IEEEtran}
\IEEEoverridecommandlockouts
\usepackage{cite}
\usepackage{amsmath,amssymb,amsfonts}
\usepackage{algorithmic}
\usepackage{graphicx}
\usepackage{textcomp}
\usepackage{xcolor}
\usepackage[T1]{fontenc}
\usepackage{helvet}

\usepackage{caption}
\usepackage{subcaption}
\def\BibTeX{{\rm B\kern-.05em{\sc i\kern-.025em b}\kern-.08em
    T\kern-.1667em\lower.7ex\hbox{E}\kern-.125emX}}
\begin{document}

\title{Leveraging PRS and PDSCH for Integrated Sensing and Communication Systems\vspace*{-1mm} \\
\thanks{This work was funded by the European Commission through the H2020 MSCA 5GSmartFact project under grant agreement number 956670.}
}

\author{Keivan Khosroshahi\IEEEauthorrefmark{1}\IEEEauthorrefmark{2}, Philippe Sehier\IEEEauthorrefmark{2}, and Sami Mekki\IEEEauthorrefmark{2},\\
\IEEEauthorrefmark{1}
Université Paris-Saclay, CNRS, CentraleSupélec, Laboratoire des Signaux et Systèmes, Gif-sur-Yvette, France\\
\IEEEauthorrefmark{2} Nokia Standards, Massy, France
\\ keivan.khosroshahi@centralesupelec.fr, \{philippe.sehier, sami.mekki\}@nokia.com
}

\maketitle
\begin{abstract}
 From the industrial standpoint on integrated sensing and communication (ISAC), the preference lies in augmenting existing infrastructure with sensing services while minimizing network changes and leveraging available resources. This paper investigates the potential of utilizing the existing infrastructure of fifth-generation (5G) new radio (NR) signals as defined by the 3rd generation partnership project (3GPP), particularly focusing on pilot signals for sensing within the ISAC framework. We propose to take advantage of the existing positioning reference signal (PRS) for sensing and the physical downlink shared channel (PDSCH) for communication, both readily available in 5G NR. However, the use of PRS for sensing poses challenges, leading to the appearance of ghost targets.
 To overcome this obstacle, we propose two innovative approaches for different PRS comb sizes within the ISAC framework, leveraging the demodulation reference signal (DMRS) within PDSCH to eliminate ghost targets. Subsequently, we formulate a resource allocation problem between PRS and PDSCH and determine the Pareto optimal point between communication and sensing without ghost targets. Through comprehensive simulation and analysis, we demonstrate that the joint exploitation of DMRS and PRS offers a promising solution for ghost target removal, while effective time and frequency resource allocation enables the achievement of Pareto optimality in ISAC.\\
\end{abstract}


\vspace*{-4mm}
\section{Introduction}
Sixth-generation (6G) wireless networks aim to revolutionize communication by integrating sensing capabilities. This paradigm shift, known as integrated sensing and communication (ISAC), promises to unlock numerous applications, from enhanced environmental monitoring to real-time object tracking and smart cities, all within a single network \cite{wild20236g}. The reference signals in fifth-generation (5G) New Radio (NR) have emerged as promising candidates for sensing tasks, owing to their ideal auto-correlation and cross-correlation properties, as well as their robust anti-noise capabilities \cite{wei20225g}. Furthermore, the utilization of reference signals in ISAC systems offers advantages such as reduced hardware expenses, simplified implementation, and decreased computational complexity \cite{ozkaptan2018ofdm}. Some recent studies have explored the potential of utilizing various reference signals for sensing applications within 5G networks. Authors in \cite{wei20225g} investigated the use of positioning reference signal (PRS) for range and Doppler estimation, comparing its performance to other 5G pilots such as demodulation reference signal (DMRS), channel state information reference signal (CSI-RS) and synchronization signal (SS). In \cite{huang2022joint}, the authors proposed a two-stage joint optimization scheme for channel estimation, target detection, and pilot selection within an ISAC framework. In \cite{wang2020multi}, a multi-range dual radar and communication system based on a pilot-based orthogonal frequency division multiplexing (OFDM) waveform is presented. CSI-RS and DMRS are used and compared in \cite{ma2022downlink} for range and velocity estimation. PRS and DMRS are jointly used in \cite{khosroshahi2024dopplerambiguityeliminationusing} to remove Doppler ambiguity in ISAC system.

Among the available pilot signals, PRS, introduced in 3rd generation partnership project (3GPP) Release 16, holds significant promise for sensing applications due to its rich time-frequency resources and flexible configuration. PRS has high resource element (RE) density and superior correlation properties compared to existing reference signals because of the staggered or diagonal RE pattern. However, PRS is not particularly designed for sensing, and its exploitation in sensing passive targets poses some challenges that cannot be ignored. More precisely, empty REs in the structure of PRS would lead to ambiguity in range estimation. Such ambiguity makes the distinction between real and ghost targets quite challenging. While the proposed solution in \cite{wei2023multiple} has focused on multiple reference signals to address range ambiguity, it is not directly applicable in ISAC scenarios due to interference between physical downlink shared channel (PDSCH) for data transmission and other reference signals such as PRS and CSI-RS. Additionally, with the proposed configuration, there will be no flexibility in choosing the comb size. Furthermore, the approach outlined in \cite{wei20225g} results in a notable decrease in the maximum detection range, especially by using high PRS comb sizes. Thus, to the best of our knowledge, a 3GPP-compliant solution to alleviate range ambiguity in the ISAC framework has not been proposed.

In this paper, we introduce two novel methods exploiting available reference signals in the 5G NR to provide sensing services for passive targets with enhanced maximum detection range without ambiguity. To enable sensing services in addition to communication, we propose an ISAC OFDM resource grid incorporating PRS for sensing and PDSCH for data transmission. To eliminate the range ambiguity, we propose two methods for different comb sizes to repurpose DMRS in PDSCH for sensing besides communication. Moreover, since sensing and communication symbols cannot share the same REs in the resource grid, we propose time and frequency resource allocation between PDSCH and PRS. Elaborating on this, we introduce and formulate a resource allocation problem between PDSCH and PRS to achieve Pareto optimality between sensing and communication without range ambiguity. Finally, through comprehensive simulations, we demonstrate the effectiveness of our methods in mitigating range ambiguity and identify the Pareto optimal point between communication and sensing.

%Leveraging an ISAC OFDM resource grid comprising PRS for radar sensing and the PDSCH for data transmission, the methods reuse the DMRS within PDSCH to remove ghost targets present in sensing with PRS while the maximum detection range for targets is extended. More precisely, the paper presents two approaches for different PRS comb sizes, facilitating simultaneous transmission of data communication and sensing symbols in an OFDM frame with separate time and frequency resource allocation. The configurations of PDSCH, DMRS, and PRS are aligned with 3GPP technical specifications for 5G New Radio (NR), ensuring compatibility with existing 5G networks. Moreover, we introduce a resource allocation problem between PDSCH and PRS and determine the Pareto optimal point between communication and sensing while effectively ghost targets are eliminated. 

\vspace*{-2mm}
\section{5G REFERENCE SIGNALs}
Based on TS 38.214 \cite{3gpp2018nr1}, the 5G-NR waveform is built on a time-frequency resource grid with the smallest time-frequency resource that can be allocated, known as physical resource block (PRB). Each PRB comprises $12$ contiguous sub-carriers and $14$ consecutive symbols. In the following two sections, we describe two of the 5G reference signals used in this work namely, PRS and DMRS.

\vspace*{-2mm}
\subsection{PRS}
\begin{figure}[!t]
    \centering
    \begin{subfigure}{0.45\columnwidth}
        \centering
        \includegraphics[width=\linewidth]{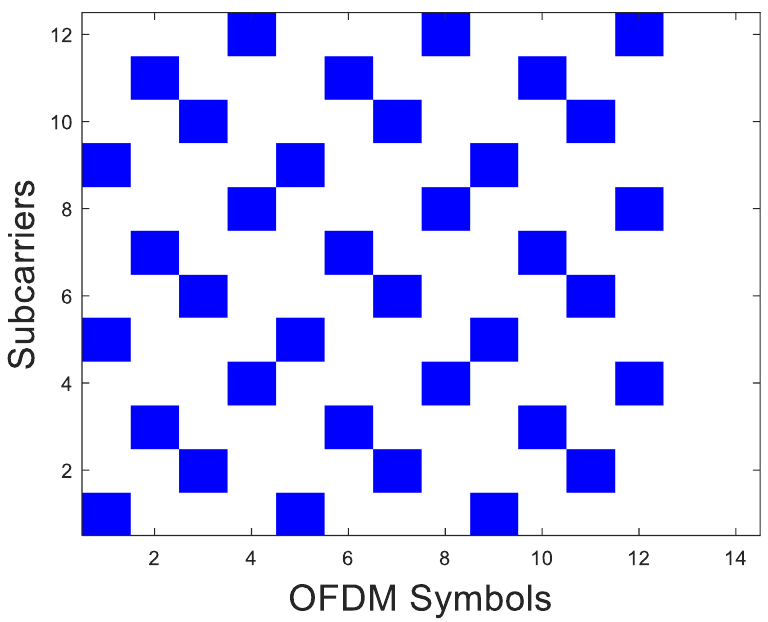}
        \caption{}
        \label{comb4}
    \end{subfigure}
    \begin{subfigure}{0.45\columnwidth}
        \centering
        \includegraphics[width=\linewidth]{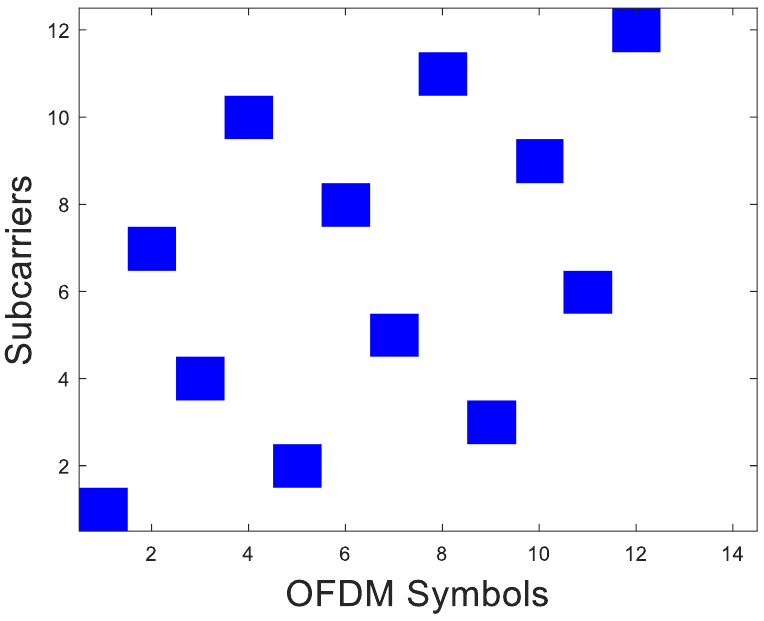}
        \caption{}
        \label{comb12}
    \end{subfigure}
    \caption{Two different PRS patterns allocation within a slot: (a) Comb size = 4, (b) Comb size = 12.}
    \label{comb}
\end{figure}

The PRS is generated as explained in TS 38.211 \cite{3gpp2018nr} as:
 \vspace*{-2mm}
\begin{equation}
   r(m) = \frac{1}{\sqrt{2}}(1 - 2c(2m)) + j\frac{1}{\sqrt{2}}(1 - 2c(2m + 1)),
   \label{r}
\end{equation}
where $c(i)$ is a Gold sequence of length-31. Generating PRS from the Gold sequence leads to suitable auto-correlation and anti-noise characteristics \cite{wei20225g}. The initial value of $c(i)$ for PRS can be found in TS 38.211 \cite{3gpp2018nr}.
Depending on the subcarrier spacing, the PRS in 5G NR can occupy a minimum of $24$ PRBs and a maximum of $272$ PRBs. PRS offers a versatile configuration of time-frequency resources to meet sensing accuracy requirements across various application scenarios. In accordance with the PRS resource mapping outlined in TS 38.211 \cite{3gpp2018nr}, PRS supports four comb patterns—Comb $2/4/6/12$—in the frequency domain and five symbol number configurations—Symbol $1/2/4/6/12$—in the time domain. 
%The structure of PRS with $12$ symbols and different comb sizes is depicted in Figs. \ref{comb}.
Figure  \ref{comb} shows two examples of comb size patterns within a slot with PRS structure of 12 symbols. 

\vspace*{-2mm}
\subsection{DMRS}
The DMRS is also generated using \eqref{r}, and the initial value of $c(i)$ for DMRS is available in \cite{3gpp2018nr}.
DMRS is generated within the designated PDSCH allocation, as outlined in TS 38.211 \cite{3gpp2018nr}. DMRS is used for channel estimation and is confined to the resource blocks (RBs) assigned for PDSCH. The structure of DMRS is designed to accommodate various deployment scenarios and use cases. The position of the DMRS symbols in the time and frequency resource grid depends on the mapping type, which can be either slot-wise (Type A) or non-slot-wise (Type B). The positions of any additional DMRS symbols are determined by a set of tables, as detailed in TS 38.211 \cite{3gpp2018nr}. Additionally, 1 to 4 OFDM symbols can be occupied by front-load DMRS in the time domain. DMRS exists within physical channels allocated for communication receivers and is transmitted using the same beam specifically chosen for communication receivers. However, we aim to reuse it for sensing purposes beside communication.

\section{ISAC System Model}
\begin{figure}[!t]
\centering
\mbox{\includegraphics[width=\linewidth]{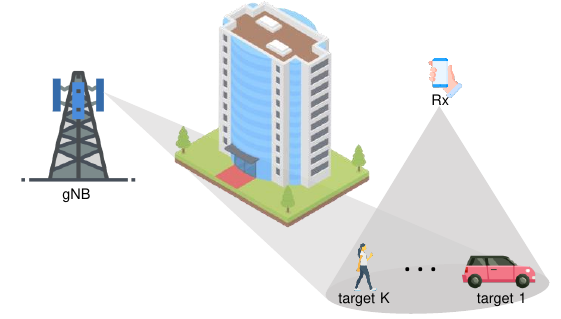}}
\caption{ISAC scenario using UE as communication and sensing receiver.}
\label{bistatic RIS2}\vspace{-5mm}
\end{figure}

The current 5G NR signals were not originally designed to support radar sensing applications. In this section, we present a methodology for effectively utilizing both communication and sensing signals within the ISAC framework. We envision a downlink ISAC system wherein a gNB transmits signals toward a sensing area containing $K$ point-like targets. The echoes from these targets are then captured by a receiver for range and velocity estimation. We assume the targets are in the line-of-sight (LoS) of the transmitter and the receiver. 
An example of the envisioned scenario is depicted in Fig. \ref{bistatic RIS2}.

The continuous time domain expression of the transmitted signal can be expressed in the following form: \cite{buzzi2020transmit}
 \vspace*{-2mm}
\begin{equation}
    s(t) = \sum_{n = 0}^{N-1}\text{rect}(\frac{t - nT_0}{T_0})\sum_{m=0}^{M-1} v(m,n) e^{j2\pi m\Delta f(t - nT_0)},
\end{equation}
where we consider an OFDM resource grid containing $N$ symbols in the time domain and $M$ sub-carriers, and $\text{rect}(t/T_0)$ represents the rectangular pulse, $\Delta f$ denotes the subcarrier spacing, and $T_0 = T_{CP} + T_s$ designates the total duration of the OFDM symbol, where $T_{s} = \frac{1}{\Delta f}$ denotes the symbol duration, and $T_{CP}$ is the cyclic prefix (CP) length. $v(m,n)$ is the complex symbol transmitted at $n$-th OFDM symbol and $m$-th sub-carrier, where $n=0,...,N-1$ and $m=0,...,M-1$ within an $M \times N$ OFDM resource grid. Some of these symbols can be set to zero or not allocated.
%For communication symbols, various modulations such as quadrature phase shift keying (QPSK),16 quadrature amplitude modulation (QAM), etc., can be utilized, and reference signals are generated as described in \eqref{r}. We should note that in one PRB and slot where $N = 14$ and $M = 12$, PRS occupies a maximum of $M_p = \frac{M}{K_{comb}}$ subcarriers and $N_p = \lfloor	\frac{N}{K_{comb}}\rfloor$ symbols where $K_{comb}$ is the comb size of PRS and $\lfloor . \rfloor$ is the floor function. In other words, the rest of the REs in one PRB and one slot are empty, \emph{i.e.}, $v(m,n) = 0$. 
The reflected signals received by the receiver can be expressed as \cite{braun2014ofdm}
 \vspace*{-2mm}
\begin{align}
    r(t)=\sum_{k=1}^K\beta_k s(t-\tau_k)e^{j2\pi f_{d,k}t} + u(t),
    \label{y_time}
\end{align}
where $\beta_k$ represents the attenuation factor of the $k$-th target, $f_{d,k}$ is the Doppler frequency shift originated from the $k$-th target, $\tau_k$ is the target $k$ delay and $u(t) \in \mathbb{C}$ is the complex additive white Gaussian noise (AWGN) with zero mean and variance of $2\sigma^2$.

Conventional OFDM receivers sample the signal at each symbol time and perform an fast fourier transform (FFT) to extract the modulated symbols. The delay and Doppler are compensated in normal conditions, and there is no inter-symbol interference. The cyclic prefix is usually exploited to remove receive timing errors and the multi-path spread. In our case, it cannot be guaranteed that the path delay and Doppler are in the usual range due to the potentially large uncertainties on the position and speed of the target. Therefore, there can be inter-symbol interference, both in the time and frequency domains. Therefore, the received samples extracted after FFT can be written as 

\vspace*{-6mm}
\begin{align}
    y(m,n) =&  \sum_{k=1}^K\beta_k e^{j2\pi n T_0 f_{d,k}} e^{-j2\pi m \Delta f \tau_k} v(m,n) + \text{ISI} \notag\\
    &+ q(m,n),
    \label{y_freq}
\end{align}
where $\text{ISI}$ indicates the inter-symbol interference term caused by Doppler and delay, $q(m,n) \in \mathbb{C}$ is the AWGN noise with zero mean and variance of $2\sigma^2$ on the $m$-th subcarrier and the $n$-th OFDM symbol obtained from sampling and FFT over $r(t)$.

\section{Estimation of the range and velocity \label{est}}
It is a well-known result that received samples undergo rotations on the time axis due to Doppler and on the frequency axis due to time shift. Estimating range and Doppler requires a two-dimensional search, for which different strategies can be considered. Several methods are described in \cite{braun2014ofdm}. We limit ourselves here to the method that requires the least computing power, which operates in 2 stages: first, the range is estimated, then the Doppler, after compensating for the delay. The time-of-flight (ToF) from the transmitter to the target and then to the receiver can be extracted from range evaluation using the periodogram \cite{pucci2021performance}. To that end, we remove the transmitted sensing symbols from the received echos by point-wise division as follows
\vspace*{-2mm}
\begin{align}
    g(m,n) =& \frac{y(m,n)}{v(m,n)},\quad &\text{If } v(m,n) \neq 0,\\
    g(m,n) =& 0,\quad &\text{If } v(m,n) = 0.
\end{align}

Then the $M$-point inverse fast fourier transform (IFFT) of the PRS on the $n$-th column of the $g(m,n)$ is calculated:

\vspace*{-5mm}
\begin{align}
    R_n(l) =& |\text{IFFT}(g(m,n))| = |\sum_{k=1}^K (\beta_k e^{j2\pi n T_0 f_{d,k}} \sum_{m=0}^{M-1} \notag \\ 
    & e^{-j2\pi m \Delta f  \frac{R^{tot}_k}{c}} e^{j2\pi\frac{ml}{M}}+ \frac{q'(m,n)}{v(m,n)}e^{j2\pi\frac{ml}{M}})|,
    \label{r_l}
\end{align}
where $l= 0,...,M-1$, $|.|$ is the absolute value and $q'(m,n) = q(m,n) + \text{ISI}$. The reason for taking the absolute value is to reduce Doppler sensitivity. We replaced $\tau_k = \frac{R^{tot}_k}{c}$ while $R^{tot}_k$ is the bistatic range, \emph{i.e.}, distance from gNB to the target and then to the receiver, $c$ is the light speed. When the arguments of $e^{-j2\pi m \Delta f  \frac{R^{tot}_k}{c}} e^{j2\pi\frac{ml}{M}}$ in \eqref{r_l} cancel each other, the maximum value occurs. The next step consists in summing the $R_n(l)$ over $N$ symbols to increase the signal-to-noise ratio (SNR):
\vspace*{-2mm}
\begin{align}
   \overline{\rm R}(l) = \frac{1}{N}\sum_{n=0}^{N-1} R_n(l).
    \label{r_l2}
\end{align}

The maximum value named $\hat{l}_k$ in \eqref{r_l2} gives the bistatic distance of each target \cite{wei20225g}: 
\vspace*{-1mm}
\begin{equation}
    \hat{R}^{tot}_k = \frac{\hat{l}_kc}{\Delta f M}.
    \label{r_tot}
\end{equation}

The range resolution can be calculated as follows
\vspace*{-1mm}
\begin{equation}
    \Delta R = \frac{c}{\Delta f M},
    \label{R_res}
\end{equation}
and the maximum detection range can be written as follows
\vspace{-1mm}
\begin{equation}
    R_{max} = \frac{c M}{\Delta f M} = \frac{c}{\Delta f}.
    \label{R_MAX}
\end{equation}

Based on \cite{li2019ofdm}, we can improve the accuracy performance by increasing the number of IFFT points $m_a$ as follows
\vspace*{-2mm}
\begin{align}
    R_n(l) =& |\text{IFFT}(g(m,n))| = |\sum_{k=1}^K (\beta_k e^{j2\pi n T_0 f_{d,k}} \sum_{m=0}^{m_aM-1} \notag \\ 
    & e^{-j2\pi m \Delta f  \frac{R^{tot}_k}{c}} e^{j2\pi\frac{ml}{m_aM}} + \frac{q'(m,n)}{v(m,n)}e^{j2\pi\frac{ml}{m_aM}})|.
    \label{r_ma}
\end{align}
In this case
\begin{equation}
    \hat{R}^{tot}_k = \frac{\hat{l}_kc}{m_a\Delta f M}. 
\end{equation}

As pointed out in \cite{li2019ofdm}, by increasing $m_a$, we can improve the range accuracy at the cost of increased computational complexity. %However, by increasing $m_a$ infinitely, ranging accuracy can only approach the Cramér-Rao lower bound (CRLB).

After compensating the delay, Doppler can be estimated as follows. We perform $N$-points FFT on the $m$-th row of the $g(m,n)$
\vspace*{-2mm}
\begin{align}
    v_m(d) =& |\text{FFT}(g(m,n))| = |\sum_{k=1}^K (\beta_k \sum_{n=0}^{N-1} e^{j2\pi n T_0 f_{d,k}} e^{-j2\pi\frac{nd}{N}}\notag \\ 
    & + \frac{q'(m,n)}{v(m,n)}e^{-j2\pi\frac{nd}{N}})|,
    \label{v_m}
\end{align}
where $d= 0,...,N-1$. Then, we perform FFT over all rows of the $g(m,n)$ and averaging over them as follows
\vspace*{-1mm}
\begin{align}
   \overline{\rm v}(d) = \frac{1}{M}\sum_{m=0}^{M-1} v_m(d).
    \label{v_m2}
\end{align}

Next, we find the index of the maximum value, \emph{i.e.}, $\hat{d}_k$ for target $k$ in \eqref{v_m2}, and the doppler can be estimated as 
\vspace*{-1mm}
\begin{equation}
    \hat{f}_{d,k} = \frac{\hat{d}_k}{T_s N}.
\end{equation}

The velocity can be obtained from Doppler as follows \cite{wei20225g} 

\begin{equation}
    v = \frac{cf_{d,k}}{2 f_c},
\end{equation}
 where $f_c$ is the carrier frequency. Hence, velocity can be estimated as
 \begin{equation}
    \hat{v} = \frac{\hat{d}_kc}{2T_s f_c N}.
\end{equation}

The velocity resolution can be written as 
\begin{equation}
    \Delta\hat{v} = \frac{c}{2T_s f_c N}, 
\end{equation}
and the maximum detectable velocity estimation is 
\begin{equation}
    \hat{v}_{max} = \frac{c}{2T_s f_c}.
    \label{v_max}
\end{equation}

We can improve the accuracy of velocity by increasing the FFT point similar to range estimation. %until the CRLB of velocity.

\subsection{Ghost Targets in Range Estimation}
PRS was originally designed for positioning. In this operating mode, time-frequency uncertainties are reduced thanks to the time-frequency servo loop. This is not the case for sensing, as the uncertainty ranges are potentially larger, and larger auto-correlation ranges of PRS need to be considered. 
PRS exhibits a recurring pattern according to the selected Comb value, 
%the periodicity of which depends on the comb value, 
as illustrated in Fig. \ref{resourcegrid} in which a comb size of $12$ is configured. %\PhS{attention, the periodicity is not visible on the figure}. 
%The empty REs in the structure of PRS can result in the appearance of ghost targets in range estimation.
%, using the algorithm discussed in section \ref{est}, while the maximum detectable range and velocity is enhanced compared to \cite{wei20225g} 
%\PhS{I do not understand this sentence above marked in comment}. 
Ghost effects make it challenging to distinguish between real and false targets, as depicted in Fig. \ref{figa:PRS} and Fig. \ref{figa:PRS2} in our simulation results. Table \ref{table3} presents the periodic distances between ghost targets, derived from the two-way propagation delay based on Eq. \eqref{R_amb}, where SCS denotes the subcarrier spacing. Notably, for one-way propagation delay, ambiguity intervals are twice the values indicated in Table \ref{table3}.
\vspace*{-1mm}
\begin{equation}
    \Delta R_{ambiguity} = \frac{c}{2 K_{comb}^{PRS}\Delta f},
    \label{R_amb}
\end{equation}
where $K_{comb}$ is the PRS comb size. Therefore, to estimate the range of the targets without encountering ambiguity, a practical approach that aligns with 3GPP standards is imperative.

%One approach to avoid the appearance of ghost targets is outlined in \cite{wei20225g}, albeit at the cost of reducing the maximum detection range by a factor of $K_{comb}$. In cases where high range resolution is critical, increasing the sub-carrier spacing, as indicated in \eqref{R_res}, becomes necessary. However, this would result in a significantly low maximum detectable range, especially with larger comb sizes. Therefore, an alternative approach to mitigate ghost targets while maintaining a high maximum detectable range is imperative, particularly for targets that are not located in the vicinity of the gNBs.
%\begin{figure}[!t]
%\centering
%\mbox{\includegraphics[width=\linewidth]{ghost_target(comb12)corp.pdf}}
%\caption{Range ambiguity which causes periodic ghost targets.}
%\label{ghost_target}
%\end{figure}

\section{Using DMRS to Remove Ghost Targets}
\begin{figure}[!t]
\centering
\mbox{\includegraphics[width=\linewidth]{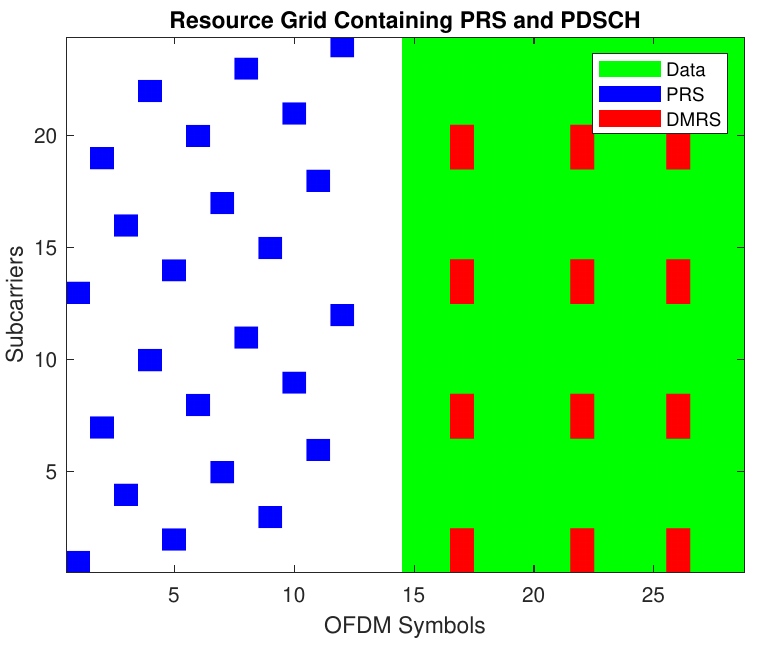}}
\caption{Resource allocation between PRS and PDSCH 
 showing 2 slots and 2 PRBs of the whole resource grid with the comb size equal to 12.}
\label{resourcegrid}
\end{figure}

The main idea involves leveraging DMRS as a reference signal available in PDSCH for ISAC use cases, where an OFDM frame contains PRS for sensing and PDSCH for communication. DMRS serves to mitigate ambiguity in range estimation and sensing. To achieve this, we generate an OFDM frame and allocate different time/frequency resources based on the application requirements for communication and sensing. Fig. \ref{resourcegrid} illustrates an example of such resource allocation in the time domain between PRS and PDSCH. DMRS not only serves communication purposes such as channel estimation but can also contribute to sensing by eliminating ghost targets without compromising accuracy and without necessitating changes to gNB configuration, as it aligns with 3GPP standards. To tackle this issue, we propose two novel algorithms for different comb sizes by allocating similar bandwidth to PDSCH and PRS.

\begin{table}[!t]
\caption{Range interval between ambiguities for regular PRS pattern (meter).}\label{table3}
\centering
\begin{tabular}{ |p{2cm}|p{1cm}|p{1cm}|p{1cm}|p{1cm}|  }
 \hline
 & comb 2 & comb 4 & comb 6 & comb 12\\
 \hline 
 SCS = 15 kHz & 4996 & 2498 & 1665 & 832\\ \hline
 SCS = 30 kHz & 2498 & 1249 & 832 & 416 \\ \hline
 SCS = 60 kHz & 1250 & 624 & 416 & 208 \\ \hline
 SCS = 120 kHz & 624 & 312 & 208 & 104\\ \hline
 SCS = 240 kHz & 312 & 156 & 104 & 52\\ \hline
\end{tabular}
\end{table}

\subsection{Algorithm 1: When Comb size is equal to $2$ or $4$}
Initially, we extract DMRS from the received slots containing PDSCH. Then, we use the range estimation algorithm explained in \ref{est} for DMRS and PRS independently. If we obtain $\overline{\rm r} $ from PRS and DMRS using \eqref{r_l2} with the same IFFT size and name them $\overline{\rm r}_{PRS}$ and $\overline{\rm r}_{DMRS}$, respectively, instead of finding the maximum value for each of them separately, we find the index of the maximum value from $\overline{\rm r}_{PRS,DMRS} = \overline{\rm r}_{PRS} \circ \overline{\rm r}_{DMRS}$, where $\circ$ is the Hadamard product. This method maintains the range resolution and maximum unambiguous range as derived in \eqref{R_res} and \eqref{R_MAX}, respectively. 
In practice, ghost targets resulting from range estimation using DMRS may appear in positions where PRS with comb sizes of $2$ and $4$ do not show any ghost targets, and vice versa. By performing element-wise multiplication of $\overline{\rm r}_{PRS}$ and $\overline{\rm r}_{DMRS}$, we effectively remove the ghost targets, enabling clear distinction between real targets and ghost ones. If PRS and PDSCH occupy different bandwidths, we can adjust the lengths of the vectors by changing the size of the IFFT points, resulting in similar vector sizes for $\overline{\rm r}_{PRS}$ and $\overline{\rm r}_{DMRS}$.

\subsection{Algorithm 2: When Comb size is equal to $6$ or $12$}
In the case that the comb size is equal to $6$ or $12$, some ghost targets of PRS and DMRS overlap by using algorithm $1$. In this case, from the extracted PRS and DMRS of the received PDSCH slots, we can obtain $g_{DMRS}(m,n)$ and $g_{PRS}(m,n)$, then add all of their columns as follows
\vspace*{-1mm}
\begin{align}
    g_{PRS}^{tot}(m) &= \sum_{n =1}^{N-1} g_{PRS}(m,n),\\
    g_{DMRS}^{tot}(m) &= \sum_{n =1}^{N'-1} g_{DMRS}(m,n),
\end{align}
where $N'$ is the number of DMRS symbols in the time domain. Then, we normalize them to balance their effect in IFFT.
\vspace*{-1mm}
\begin{align}
    \Tilde{g}_{PRS}^{tot}(m) &= \frac{g_{PRS}^{tot}(m)}{\text{max}_{m}\{g_{PRS}^{tot}(m)\}},\\
    \Tilde{g}_{DMRS}^{tot}(m) &= \frac{g_{DMRS}^{tot}(m)}{\text{max}_{m}\{g_{DMRS}^{tot}(m)\}}.
\end{align}

Afterwards, we perform $M$-point IFFT on the summation of $\Tilde{g}_{PRS}^{tot}(m)$ and $\Tilde{g}_{DMRS}^{tot}(m)$ as follows
\vspace*{0mm}
\begin{equation}
    r(l) = |\text{IFFT}(\Tilde{g}_{PRS}^{tot}(m) + \Tilde{g}_{DMRS}^{tot}(m))|.
\end{equation}

Eventually, we can find the index of the maximum value of $r(l)$ to estimate the bistatic range of targets. The rest of the parameters, \emph{i.e.}, bistatic distance, range resolution, and the maximum detection range can be calculated using \eqref{r_tot}, \eqref{R_res}, and \eqref{R_MAX}, respectively.
In this way, we are actually disturbing the periodicity of the pattern of PRS and DMRS, which leads to suppression of the magnitude of the fake peaks. Therefore, we can easily differentiate between real and fake targets. 
If PRS and PDSCH have different bandwidths, we pad zeros to the reference signal with the smaller bandwidth so eventually, $\Tilde{g}_{PRS}^{tot}(m)$ and $\Tilde{g}_{DMRS}^{tot}(m)$ can have similar vector sizes.

\section{Resource Allocation to reach Pareto optimality in ISAC}
In this section, we define an optimization problem for resource allocation between PDSCH and PRS to find the Pareto optimal point between communication and sensing without ghost targets through the two proposed algorithms. To that end, we define the maximum throughput of the received PDSCH per one slot and one PRB as $R_0$, $\frac{c}{\Delta f}$ as the range resolution metric, and $\frac{c}{2T_s f_c}$ as the velocity resolution metric. We introduce the following resource allocation problem to find the Pareto optimal point. 

\vspace*{-3mm}
\begin{subequations}
\begin{align}
 \underset{m_0,n_0,...,m_K,n_K }
{\text{max}}&
 \frac{\alpha_0 m_0 n_0}{R_{\text{max}}} R_0 -  \sum_{k=1 }^{K}( \frac{\gamma_{k,1} }{d_{\text{max}} m_k} \frac{c}{\Delta f} \notag \\
&+ \frac{\gamma_{k,2} }{v_{\text{max}} n_k} \frac{c}{2T_s f_c} ),  \label{st.0}\\
\textrm{s.t. }&m_i \in \{1,...,M_{\text{max}}-1\}, \textrm{  } i = 0,...,K, \\
&  n_i \in \{1,...,N_{\text{max}}-1\} ,\textrm{  } i= 0,...,K,  \\
& \sum_{i=0}^K m_i = M_{\text{max}}, \\
& \sum_{i=0}^K n_i = N_{\text{max}}, 
%& \sum_{i=1}^K m_i \leq m_0,\label{st.5}\\
%& \sum_{i=1}^K n_i \leq n_0,\label{st.6}\\
\end{align}
\end{subequations}
where $\alpha_0$, $\gamma_{k,1}$, and $\gamma_{k,2}$ are the communication weight, the range and velocity resolution weights of the target $k$, respectively, which can be chosen based on the priorities of the application and $\alpha_0 + \sum_{k=1}^K (\gamma_{k,1} +\gamma_{k,2}) = 1 $. $R_{\text{max}}$, $d_{\text{max}}$, and $v_{\text{max}}$ are the maximum throughput, range, and velocity resolution that can be achieved by allocating all the available resources in the time, \emph{i.e.}, $N_{\text{max}}$, and frequency, \emph{i.e.}, $M_{\text{max}}$, domain for communication data transmission, range and velocity estimation, respectively. $m_0$ and $n_0$ are the number of PRBs and slots, respectively, allocated for communication, and $m_k$ and $n_k$ are the number of PRBs and slots, respectively, allocated for the range and velocity estimation of the target $k$. By solving the introduced optimization problem, we can find the Pareto optimal point of resource allocation between communication and sensing.
\vspace*{-1mm}
\begin{figure}[!t]
    \centering
    \begin{subfigure}{\columnwidth}
        \centering
        \includegraphics[width=\linewidth]{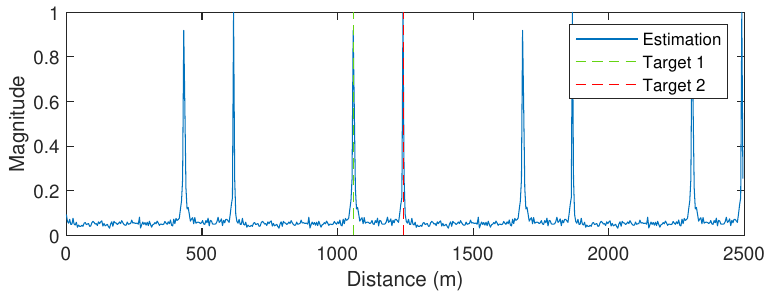}
        \vspace{-0.5cm}
        \caption{Range estimation with PRS only.}
         \vspace{0.3cm}
        \label{figa:PRS}
    \end{subfigure}
    %\vspace{-1mm}
    \begin{subfigure}{\columnwidth}
        \centering
        \includegraphics[width=\linewidth]{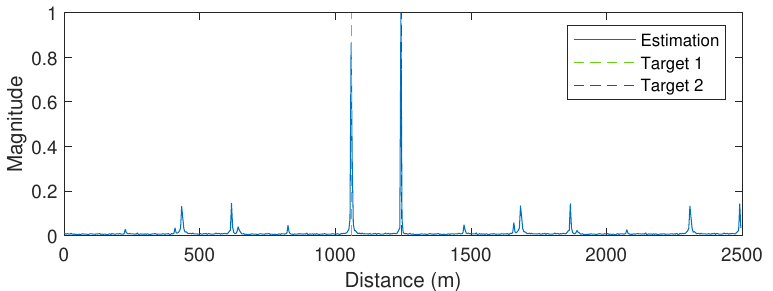}
        \caption{Range estimation with PRS and DMRS combined.}
        \label{figb:PRS_DMRS}
    \end{subfigure}
     \caption{Ghost target elimination with the presence of 2 targets with $K_{comb} = 4$,  $f_c=28$GHz and SCS=$120$kHz.}
    \label{fig:PRSDMRcomb4}
\end{figure}

%\begin{figure}[!t]
%\centering
%\mbox{\includegraphics[trim=0 0 \linewidth 0.5\linewidth,clip,width=\linewidth]{PRS_PRSDMRS_COMB4_2target_crop.pdf}}
%\caption{Ghost target elimination with the presence of 2 targets with $K_{comb} = 4$,  $f_c=28$GHz and SCS=$120$kHz.}
%\label{PRSDMRS_comb4_2target}
%\end{figure}

\section{Simulation Results}
\begin{figure}[!t]
    \centering
    \begin{subfigure}{\columnwidth}
        \centering
        \includegraphics[width=\linewidth]{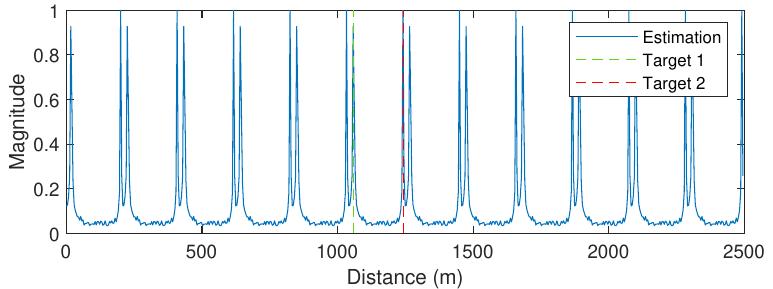}
        \vspace{-0.5cm}
        \caption{Range estimation with PRS only.}
         \vspace{0.3cm}
        \label{figa:PRS2}
    \end{subfigure}
    %\vspace{2cm}
    \begin{subfigure}{\columnwidth}
        \centering
        \includegraphics[width=\linewidth]{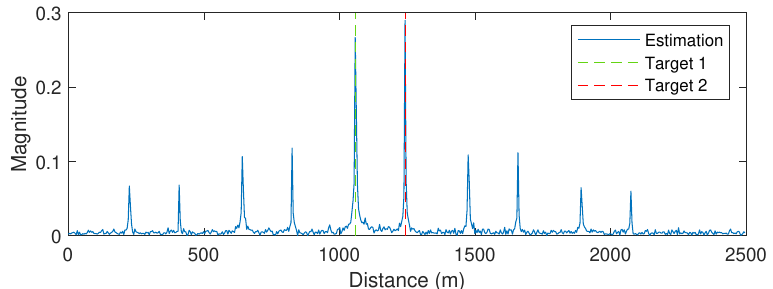}
        \caption{Range estimation with PRS and DMRS combined.}
        \label{figb:PRS_DMRS2}
    \end{subfigure}
    \caption{Ghost target elimination with the presence of 2 targets with $K_{comb} = 12$,  $f_c=28$GHz and SCS=$120$kHz.}
    \label{fig:PRSDMRcomb12}
\end{figure}

To prove our method, we utilize Matlab 5G toolboxes, making this work compatible with 3GPP standards. We simulated a scenario in which two targets with the bistatic distance of $1057$m and $1242$m and the velocity of $5$m/s for both are considered. SNR should be enough to be able to detect PRS and DMRS, $f_c=28$GHz and SCS equals to $120$kHz. PRS number of symbols is $12$, DMRS configuration type is $2$, there are $3$ DMRS symbols placed at $\{3,8,12\}$ symbols of each slot, PDSCH mapping type is "A", the code rate used to calculate transport block sizes is $490/1024$ based on \cite{3gpp2018nr1}. Virtual resource block (VRB) bundle size is $4$, and VRB to PRB interleaving is disabled. $\beta_1 =\beta_2=1$ and we consider the 16 quadrature amplitude modulation (16-QAM) modulation for PDSCH. 
In Fig. \ref{figa:PRS}, we can see the presence of multiple ghost targets for range estimation of passive targets when using only PRS with the comb size equal to $4$. However, in Fig. \ref{figb:PRS_DMRS}, the combination of PRS and DMRS can eliminate the range ambiguity, as explained in algorithm 1. It is worth mentioning that Fig. \ref{figa:PRS} and \ref{figb:PRS_DMRS} are normalized. Figures \ref{figa:PRS2} and \ref{figb:PRS_DMRS2} show the ability of algorithm $2$ to remove the ghost targets when the PRS comb size is equal to $12$.

%\begin{figure}[!t]
%\centering
%\mbox{\includegraphics[width=\linewidth]{PRS_PRSDMRS_COMB12_2target_corp.pdf}}
%\caption{Ghost target elimination with the presence of 2 targets with $K_{comb} = 12$, $f_c=28$GHz %and SCS=$120$kHz.}
%\label{PRSDMRS_comb12_2target}\vspace{-5mm}
%\end{figure}

\begin{figure}[!t]
\centering
\mbox{\includegraphics[width=\linewidth]{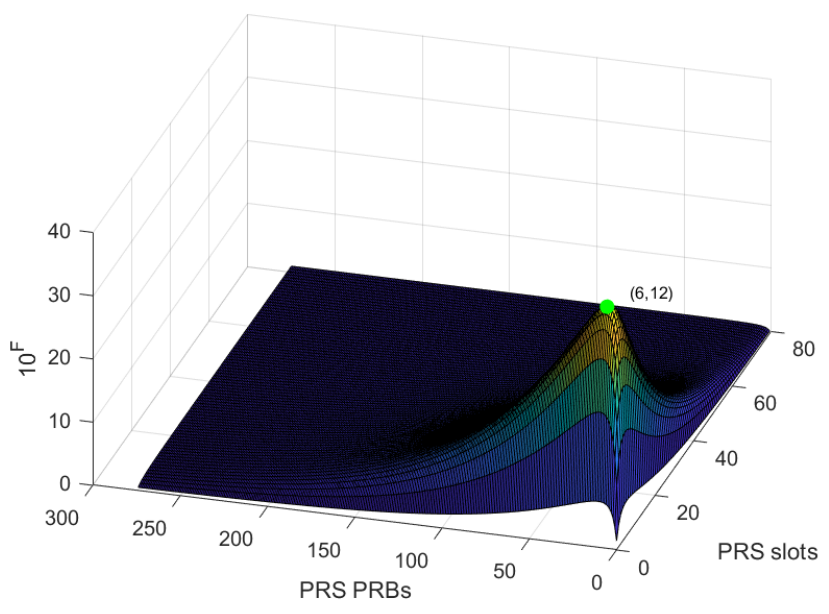}}
\caption{Pareto optimality in resource allocation between communication and sensing.}
\label{RS}\vspace{-5mm}
\end{figure}

Once the ghost targets are removed, we can obtain the Pareto optimal point from the introduced resource allocation optimization, as illustrated in Fig. \ref{RS}, where $F$ is the objective function of the optimization in \eqref{st.0}. In this simulation, we considered the maximum available PRBs and slots in 5G NR for PRS, which are $272$ and $80$, respectively. We also set $K=1$ and $\alpha_0 = 2$, $\gamma_{1,1} = \gamma_{1,2} = 1$. As depicted in Fig. \ref{RS}, with such configurations and weights, the Pareto optimal point is achieved by allocating $6$ slots and $12$ PRBs for PRS and the rest to PDSCH.

\section{Conclusion}
In this work, we tackled the ghost targets issue of PRS in the ISAC framework by leveraging DMRS available in PDSCH for passive target localization. We proposed two methods for different comb sizes, fully compliant with 3GPP standards and readily implementable in current 5G networks, distinguishing our approach from prior works. Moreover, our solution facilitates sensing without introducing additional overhead, necessitating changes in gNB configurations, or deteriorating positioning accuracy while the maximum detectable range is enhanced. Finally, we introduced a resource allocation problem between PDSCH and PRS to determine the Pareto optimal point between communication and sensing while effectively eliminating ghost targets using the proposed methods.

\bibliographystyle{IEEEtran}
\bibliography{ref}

\end{document}